\documentclass[preprint2]{aastex6}
\usepackage{apjfonts}

\usepackage{multirow}

\gdef\flux_radius{\textsc{flux\_radius}}

\gdef\mum{$\mu\mathrm{m}$}
\gdef\24mum{$24\,\mu\mathrm{m}$}
\gdef\arcsec{^{\prime\prime}}

\gdef\HAWKI{\mbox{HAWK-I}}

\shortauthors{Brammer et al.}
\shorttitle{$K_s$-band imaging of the Frontier Fields}
\slugcomment{Draft (\today)}

\begin{document}

\title{Ultra-deep $K_s$-band Imaging of the \textit{Hubble} Frontier Fields}


\author{Gabriel B. Brammer\altaffilmark{1},
Danilo Marchesini\altaffilmark{2},
Ivo Labb\'e\altaffilmark{3}, 
Lee Spitler\altaffilmark{4,5}, 
Daniel Lange-Vagle\altaffilmark{1,2},
Elizbeth A. Barker\altaffilmark{1},
Masayuki Tanaka\altaffilmark{6},
Adriano Fontana\altaffilmark{7},
Audrey Galametz\altaffilmark{8},
Anna Ferr\'e-Mateu\altaffilmark{9},
Tadayuki Kodama\altaffilmark{6},
Britt Lundgren\altaffilmark{10},
Nicholas Martis\altaffilmark{2},
Adam Muzzin\altaffilmark{11},
Mauro Stefanon\altaffilmark{3},
Sune Toft\altaffilmark{12},
Arjen van der Wel\altaffilmark{13},
Benedetta Vulcani\altaffilmark{14}, 
Katherine E. Whitaker\altaffilmark{15,16}
}

\email{brammer@stsci.edu}

\altaffiltext{1}{Space Telescope Science Institute, 3700 San Martin Dr., Baltimore, MD 21218, USA}
\altaffiltext{2}{Physics and Astronomy Department, Tufts University, Medford, MA 02155, USA}
\altaffiltext{3}{Leiden Observatory, Leiden University, NL-2300 RA Leiden, Netherlands}
\altaffiltext{4}{Department of Physics \& Astronomy, Macquarie University, Sydney, NSW 2109, Australia}
\altaffiltext{5}{Australian Astronomical Observatories, PO Box 915 North Ryde NSW 1670, Australia}
\altaffiltext{6}{National Astronomical Observatory of Japan 2-21-1 Osawa, Mitaka, Tokyo 181-8588, Japan}
\altaffiltext{7}{INAF, Osservatorio Astronomico di Roma, Via Frascati 33, I-00040, Monteporzio, Italy}
\altaffiltext{8}{Max-Planck-Institut fur extraterrestrische Physik (MPE), Giessenbachstr. 1, D-85748 Garching, Germany}
\altaffiltext{9}{Subaru Telescope, 650 N. Aohoku Pl, 96720 Hilo, HI, USA}
\altaffiltext{10}{Department of Astronomy, University of Wisconsin, Madison, WI 53706, USA}
\altaffiltext{11}{Kavli Institute for Cosmology, Cambridge University, Madingley Rd, Cambridge, UK, CB3 0HA}
\altaffiltext{12}{Dark Cosmology Centre, Niels Bohr Institute, University of Copenhagen, Juliane Maries Vej 30, DK-2100 Copenhagen, Denmark}
\altaffiltext{13}{Max Planck Institute for Astronomy (MPIA), K\"onigstuhl 17, 69117, Heidelberg, Germany}
\altaffiltext{14}{School of Physics, Tin Alley, University of Melbourne VIC 3010, Australia}
\altaffiltext{15}{Department of Astronomy, University of Massachusetts Amherst, Amherst, MA 01003, USA}
\altaffiltext{16}{Hubble Fellow}


\begin{abstract}

We present an overview of the ``KIFF'' project, which provides ultra-deep $K_s$-band imaging of all six of the \textit{Hubble} Frontier Fields clusters Abell 2744, MACS-0416, Abell S1063, Abell 370, MACS-0717 and MACS-1149.  All of these fields have recently been observed with large allocations of Directors' Discretionary Time with the \textit{HST} and \textit{Spitzer} telescopes covering $0.4 < \lambda < 1.6$\,\mum\ and 3.6--4.5\,\mum, respectively.  \textit{VLT}/\HAWKI\ integrations of the first four fields reach 5$\sigma$ limiting depths of $K_s\sim26.0$ (AB, point sources) and have excellent image quality (FWHM$\sim0\farcs4$).  Shorter \textit{Keck}/MOSFIRE integrations of the MACS-0717 (MACS-1149) field better observable in the north reach limiting depths $K_s$=25.5 (25.1) with seeing FWHM$\sim$$0\farcs4$ ($0\farcs5$).  In all cases the $K_s$-band mosaics cover the primary cluster and parallel \textit{HST}/ACS+WFC3 fields.  The total area of the $K_s$-band coverage is 490~arcmin$^2$.  The $K_s$-band at 2.2\,\mum\ crucially fills the gap between the reddest HST filter (1.6\,\mum\,$\sim H$~band) and the IRAC 3.6\,\mum\ passband.  While reaching the full depths of the space-based imaging is not currently feasible from the ground, the deep $K_s$-band images provide important constraints on both the redshifts and the stellar population properties of galaxies extending well below the characteristic stellar mass across most of the age of the universe, down to, and including, the redshifts of the targeted galaxy clusters ($z \lesssim 0.5$).  Reduced, aligned mosaics of all six survey fields are provided accompanying this manuscript.

\end{abstract}


\section{Introduction}
\label{s:introduction}

During the last two decades, near-infrared (NIR) imaging has gained a dominant role in studies of galaxy formation and evolution, enabling transformational advances in our understanding of galaxy populations at early cosmic times.

Detection of galaxies in the $K$-band ($\lambda_{\rm eff} \sim 2.2$~$\mu$m) provided the first opportunity to construct a comprehensive picture of the population of galaxies in the early universe, as it enabled the discovery of galaxies at $z > 2$ that are faint at observed optical (rest-frame ultraviolet) wavelengths due to evolved stellar populations and/or significant amount of dust extinction \citep[e.g.,][]{franx:03, labbe:03, forster:04, minowa:05, kajisawa:06, brammer:07, taylor:09a}. In fact, these galaxies, which dominate the high-mass end of the high-$z$ galaxy population, were previously missed by rest-frame UV selection techniques (e.g., $U$-dropout galaxies) \citep[e.g.,][]{vandokkum:06}. Imaging in the $K$-band allows for the direct sampling of rest-frame wavelengths longer than the Balmer break out to $z\approx5$. Sampling the rest-frame optical wavelength regime is critical for high-$z$ studies, as it is significantly less affected by dust obscuration and a better probe of the galaxy stellar mass compared to the rest-frame UV, which is more sensitive to unobscured star formation \citep[e.g., ][]{fontana:06, marchesini:09}. 

While imaging at even longer wavelengths ($\lambda>3$~$\mu$m) is needed to probe the rest-frame optical emission of galaxies in the first billion years of cosmic history, deep sub-arcsecond resolution imaging at these wavelengths will not be available until the launch of the {\it James Webb Space Telescope} ({\it JWST}). Deep $K$-band data with good  (i.e., FWHM~$<1\arcsec$) image quality have been fundamental to fully exploit {\it Spitzer}-IRAC imaging data characterized by much lower spatial resolution, allowing for the mitigation of the blending of sources in IRAC images \citep[e.g.,][]{labbe:05}. Finally, for space-based studies, the $K$ band fills the gap between the reddest {\it Hubble Space Telescope} ({\it HST}) filter (i.e., F160W, with $\lambda \sim 1.6$~$\mu$m) and the IRAC 3.6~$\mu$m passband, greatly improving the constraints on both the photometric redshifts and the inferred stellar-population properties of galaxies in the survey areas. 

It is not surprising that, given the aforementioned reasons, all successful extragalactic surveys performed in the last fifteen years to investigate galaxies populations in the early universe invested significant resources to obtain deep NIR imaging data with good image quality, especially in the $K$ band. Indeed, our understanding of galaxy formation and evolution in the recent years has impressively proceeded forward mostly driven by NIR surveys progressively deeper, wider, and with better image quality, such as FIRES \citep{labbe:03}, Subaru Super Deep Field (AO) \citep{minowa:05}, MUSYC \citep{quadri:07, taylor:09b}, FIREWORKS \citep{wuyts:fireworks}, MODS \citep{kajisawa:11}, NMBS \citep{whitaker:nmbs}, UltraVISTA \citep{mccracken:ultravista}, TENIS \citep{hsieh:tenis}, WIRDS \citep{bielby:12}, ZFOURGE \citep{spitler:12}, and HUGS \citep{hugs}.

The latest effort to further our knowledge of galaxy formation and evolution is represented by the {\it HST} Frontier Fields (HFF) program \citep{lotz:16}. The HFF program is a multi-cycle \textit{Hubble} program consisting of 840 orbits of Director's Discretionary (DD) time that is imaging six deep fields centered on strong lensing galaxy clusters in parallel with six deep blank fields. The primary science goals of the twelve HFF fields are to 1) reveal the population of galaxies at $z=$5--10 that are 10--50 times fainter intrinsically than any presently known, 2) solidify our understanding of the stellar masses and star formation histories of faint galaxies, 3) provide the first statistically meaningful morphological characterization of star-forming galaxies at $z>5$, and 4) find $z>8$ galaxies magnified by the cluster lensing, with some bright enough to make them accessible to spectroscopic follow-up. Along with {\it HST}, the {\it Spitzer Space Telescope} has devoted 1000 hours of DD time to image the HFF fields at 3.6$\mu$m and 4.5$\mu$m with IRAC (Capak et al., in prep).  

The Frontier Fields initiative is complemented by a number of separate supporting general observer programs, such as deep \textit{HST} ultraviolet imaging (Siana et al., in prep) and grism spectroscopy \citep[GLASS, ][]{treu:glass}, and deep far-infrared imaging with \textit{Herschel} \citep{rawle:16}.  Whereas the main goal of the HFF is to explore the galaxy population in the first billion years of cosmic history, this dataset is also unique for its combination of surveyed area, multiwavelength coverage and depth for studies of galaxy evolution across most of the age of the universe, down to, and including, the redshifts of the targeted galaxy clusters ($z\approx$~0.3--0.5). 

The space-based HFF data alone, however, are not sufficient to robustly characterize red galaxies at $z\gtrsim3$ because the WFC3/IR $H_{160}$ band lies on the UV side of the rest-frame optical Balmer/4000\AA~break at these redshifts, resulting in sub-optimal accuracies in the photometric redshifts and stellar population properties \citep[e.g., stellar mass and rest-frame optical color;][]{muzzin:09}. Very deep $K$-band imaging is required to improve the precision of both photometric redshifts and derived stellar population properties (see \S\ref{s:discussion}). Moreover, at $z>$8--9, the K-band data helps to constrain the Lyman-break redshifts (e.g., \citealt{bouwens:13}) and increases the wavelength lever arm for measuring the redshift evolution of the rest-frame UV slopes (i.e., dust content and/or metallicity) of the first galaxies (\citealt{bouwens:12a}; \citealt{bouwens:13}).

To resolve this issue, i.e., the lack of $K$-band data over the HFF, we executed a program---``KIFF'': ``$K$-band Imaging of the Frontier Fields''---to image all twelve HFF pointings in the $K_s$ band down to a comparable depth of the {\it HST} data using the instruments HAWK-I (mounted on the {\it VLT}) and MOSFIRE (mounted on the {\it Keck I} telescope). This paper describes the observations and analysis of $K_s$ band data obtained by the KIFF program and is accompanied by the first full public release of the reduced full-depth $K_s$ band imaging mosaics. This paper is structured as follows. The observations and data reduction are presented in \S\ref{s:observations}, while \S\ref{s:mosaic} presents the drizzled $K_s$ mosaics, along with the quantification of the image quality, the noise properties and depth, and the completeness. Finally, \S\ref{s:discussion} concludes showing a few crucial improvements enabled by the deep $K_s$ data in consort with the space-based \textit{HST} and \textit{Spitzer} Frontier Fields imaging.  AB magnitudes are used throughout, with $m_{K, \mathrm{AB}} - m_{K, \mathrm{Vega}} = 1.826$ and 1.821 and pivot wavelengths \citep{tokunaga:05} of 2.152~\mum\ and 2.147~\mum\ for the \HAWKI\ and MOSFIRE $K_s$ filter bandpasses, respectively.


\begin{deluxetable*}{clrrclcrrr}
\tabletypesize{\small}
\tablewidth{0pt} 
\tablecaption{Field summary \label{tab:summary}} 
\tablehead{ 
\colhead{Field} & \colhead{$z$\tablenotemark{a}} & \colhead{R.A.} & \colhead{Dec.} & \colhead{Instrument} & \colhead{Epoch} & \textit{HST}\tablenotemark{b} & \colhead{$t$, hours} & {Depth\tablenotemark{c}} & \colhead{FWHM}}
\startdata 
Abell 2744   & 0.31 & 00:14:21.2 & $-$30:23:50 & \textit{VLT}/\HAWKI     & 2013 Oct 24--2013 Dec 24 &  C+P & 29.3 &  26.0 & $0\farcs39$ \\ 
MACS-0416    & 0.40 & 04:16:08.9 & $-$24:04:28 & \textit{VLT}/\HAWKI     & 2013 Oct 25--2014 Feb 23 &  C+P & 25.8 &  26.0 & $0\farcs36$ \\ 
Abell S1063  & 0.35 & 22:49:01.1 & $-$44:32:13 & \textit{VLT}/\HAWKI     & 2015 Jul 8--Sep 22  &  C+P & 27.9 &  26.0 & $0\farcs39$ \\ 
Abell 370    & 0.38 & 02:40:03.3 & $-$01:36:23 & \textit{VLT}/\HAWKI     & 2015 Jul 26--2016 Jan 28 & C+P & 28.3 &  26.0 & $0\farcs35$ \\ 
\vspace {-0.2cm}\\ 
\multirow{2}{*}{MACS-0717}    & \multirow{2}{*}{0.55} & \multirow{2}{*}{07:17:34.0} & \multirow{2}{*}{$+$37:44:49} &  \multirow{2}{*}{\textit{Keck}/MOSFIRE}  &  \multirow{2}{*}{2015 Jan 26/27, 2016 Jan 21/22}  & C & 4.3 & 25.3 & $0\farcs42$ \\ 
        &                      &                             &                             &                                          &      & P & 3.8 & 25.5 & $0\farcs49$ \\ 
\vspace {-0.2cm}\\ 
\multirow{4}{*}{MACS-1149}    & \multirow{4}{*}{0.54} & \multirow{4}{*}{11:49:36.3} & \multirow{4}{*}{$+$22:23:58} & \multirow{2}{*}{\textit{Keck}/MOSFIRE}  & \multirow{2}{*}{2015 Feb 24, 2016 Jan 21/22}      & C & 5.5 & 25.2 & $0\farcs53$ \\
&  & &         &      &   & P & 4.8 & 25.1 & $0\farcs54$ \\
\vspace {-0.4cm}\\ 
&  & &         & \textit{VLT}/\HAWKI\tablenotemark{d}     & 2013 Mar 21--2014 Jun 9      & C & 5.3 & 25.0 & $0\farcs41$ \\
\enddata 
\tablenotetext{a}{Cluster redshift.}
\tablenotetext{b}{Coverage of \textit{Hubble} survey fields: C=Cluster, P=Parallel.  Most \HAWKI\ pointings cover the two \textit{HST} fields simultaneously (C+P; Fig.~\ref{fig:layout}). \\The MOSFIRE observations of MACS-0717 and MACS-1149 require two separate pointings of the instrument to cover the cluster and pa-\\rallel \textit{HST} fields, which have the different characteristics as indicated.}
\tablenotetext{c}{Depth is defined as the 5$\sigma$ limiting magnitude for point sources, measured in $D=0\farcs6$ apertures (\S\ref{s:depth}).} 
\tablenotetext{d}{Archival observations from ESO program \href{http://archive.eso.org/wdb/wdb/eso/eso_archive_main/query?prog_id=090.A-0458(A)}{090.A-0458} (PI: Infante).}
\end{deluxetable*}

\section{Observations \& Data Reduction}
\label{s:observations}

Table~\ref{tab:summary} provides a summary of the survey fields and the characteristics of the $K_s$-band observational program.  Additional details on the observing strategy and image reduction procedures are provided in the subsections below.

\subsection{Observation log}
\label{s:log}

\begin{figure*}[!t]
\epsscale{1.17}
\plotone{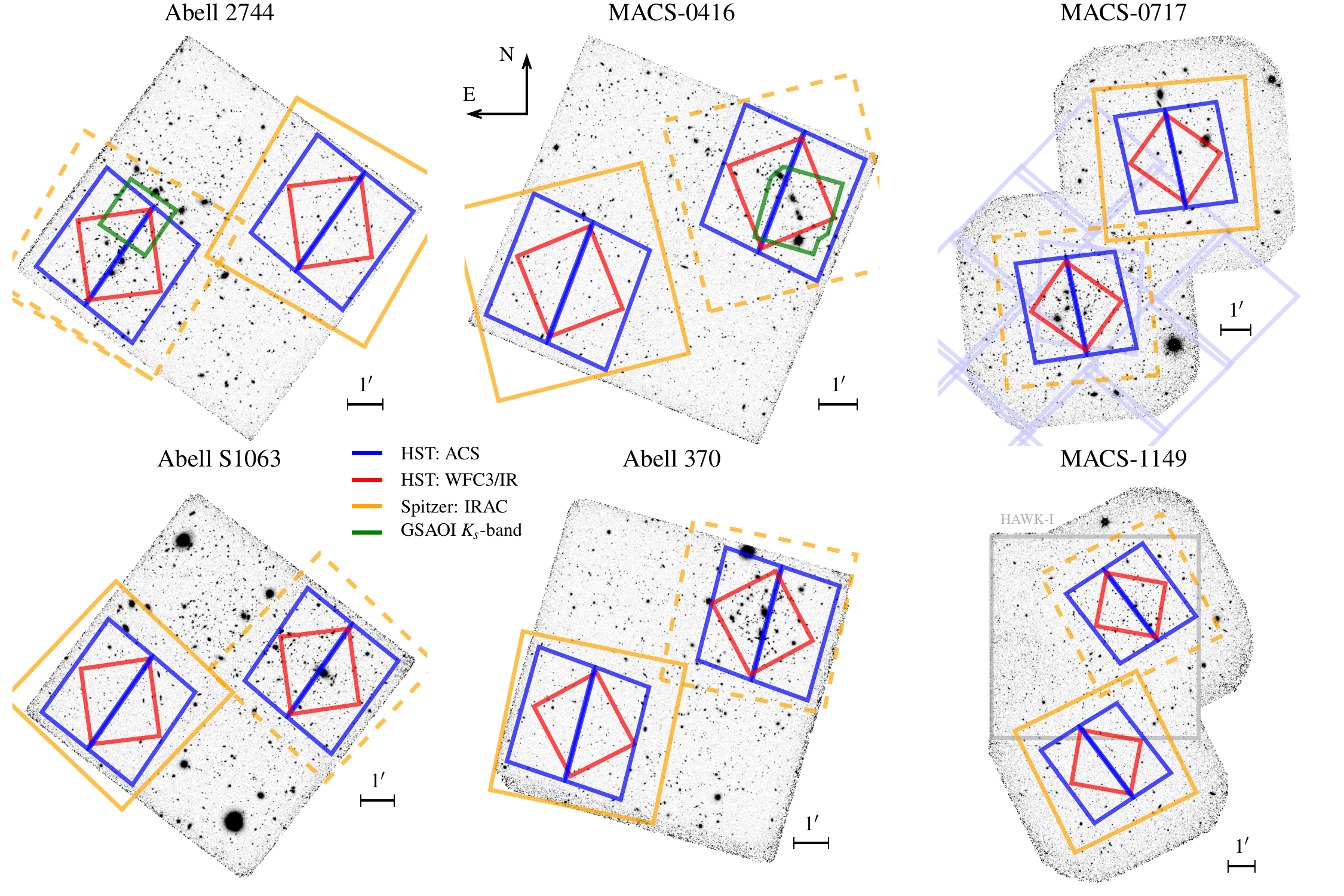} 
\caption{Layout of the Frontier Fields $K_s$-band mosaics.  The positions of the \textit{HST} cluster and parallel fields are shown in the blue (ACS optical) and red (WFC3 IR) polygons.  The light blue polygons in the MACS-0717 field show additional wide-field ACS imaging coverage from  programs GO-9722 and GO-10420 \citep[PI: Ebeling;][]{ma:11}.  The area covered by deep imaging in the \textit{Spitzer}/IRAC 3.6 and 4.5\,\mum\ channels is shown in orange, with the cluster fields indicated by the dashed lines.  The \HAWKI\ field of view is perfectly suited (Abell 2744, MACS-0416, Abell S1063 and Abell 370) for simultaneous imaging of the cluster+parallel field pairs, which require two separate pointings with MOSFIRE (MACS-0717 and MACS-1149).  The footprints of the AO-assisted GSAOI $K_s$-band imaging of the Abell 2744 and MACS-0416 fields \citep{schirmer:15} are shown in green and the footprint of additional archival \HAWKI\ coverage of the MACS-1149 field is indicated by the gray square in the lower right panel.\label{fig:layout}}  
\end{figure*}

$K_s$-band observations of the Abell 2744 and MACS-0416 fields were obtained between 2013 Oct and 2014 Feb with the High Acuity Wide-field K-band Imager \citep[\HAWKI;][]{hawki, hawki2} on the 8.2\,m UT4 telescope at the ESO \textit{Very Large Telescope} (ESO program \mbox{092.A-0472}).  \HAWKI\ $K_s$-band observations of the Abell S1063 and Abell 370 fields were obtained between 2015 Jul and 2016 Jan from a subsequent program (095.A-0533).  \HAWKI\ is composed of four chips with 2048$\times$2048 $0\farcs106$ pixels, and the full $7\farcm5\times7\farcm5$ \HAWKI\ field-of-view is perfectly suited to cover both the primary and parallel \textit{HST} ACS optical and WFC3 IR fields simultaneously in a single pointing (see Fig.\ref{fig:layout}).

The \HAWKI\ observations were divided into individual service mode Observation Blocks (OBs); typically one or two OBs of a given field were observed per night as conditions allowed, and occasionally multiple fields were observed on the same night.  The execution of each OB lasted either 60 or 90 minutes, depending on how many exposures were obtained in the block.  The exposures were roughly one minute each, with multiple coadds of shorter reads making up the exposure ($\mathrm{NDIT}\times\mathrm{DIT} = 4\times15$\,s~=~60\,s for Abell 2744, 7$\times$8\,s~=~56\,s for MACS-0416, 8$\times$12\,s~=~96\,s for Abell S1063 and 8$\times$12\,s~=~96\,s for Abell 370\footnote{Longer DITs are preferable to reduce instrument overheads; the shorter DITs on the MACS-0416 field were done do accommodate \textit{VLT} service mode restrictions on bright source saturation, which have been eased since ESO Period 93.}).  The telescope was offset with random dithers between each exposure to facilitate sky subtraction;  observations taken before 2013 Nov 4 were dithered within a 20$^{\prime\prime}$ box, which was subsequently increased to 40$^{\prime\prime}$ to improve sky subtraction in the cluster cores crowded with bright galaxies and intra-cluster light.  The total on-sky integration times for the Abell 2744, MACS-0416, Abell S1063 and Abell 370 fields are 29.3, 25.8, 27.9, and 28.3 hours, respectively.  This includes some time taken with unfavorable image quality or transparency conditions in service mode;  all on-sky exposures are included in the final mosaics with relative weighting designed to favor the optimal observing conditions (see \S\ref{s:mosaic} and Eq.~\ref{eq:weight}).

Imaging observations of the MACS-0717 and MACS-1149 fields in the $K_s$ filter were obtained on 2015 Jan 26/27 and 2015 Feb 24 (program N097M), respectively, and both fields again on 2016 Jan 21/22 (program N135M), with the Multi-Object Spectrometer For Infra-Red Exploration \citep[MOSFIRE;][]{mosfire} on the 10\,m \mbox{\textit{Keck I}} telescope.  The MOSFIRE detector consists of $2048\times2048$ $0\farcs1798$ pixels; the resulting $6\farcm1\times6\farcm1$ field-of-view is unfortunately slightly too small to cover both \textit{HST} fields simultaneously and therefore requires two pointings to cover the entire deep \textit{HST} Frontier Fields imaging area (Fig.~\ref{fig:layout}).  MOSFIRE exposures were obtained in a $3\times3$ ``Box9'' dither pattern spaced roughly 40$^{\prime\prime}$ between offset positions.  The total integration times on the MOSFIRE survey fields are summarized in Table~\ref{tab:summary}.  The individual detector integration times (DIT) were adjusted on the fly during the observing run to keep the total counts within the linear regime of the detector, while maintaining $\mathrm{NDIT}\times\mathrm{DIT} \sim 40$\,s.

Additional archival \HAWKI\ imaging of the MACS-1149 fields was obtained from the program 090.A-0458.  These data were obtained with one \HAWKI\ chip centered on the $z=9.7$ candidate from \cite{zheng:12}; they largely cover the HFF cluster field but the pointing was not optimized to also include the parallel field whose location was defined later (see Fig.~\ref{fig:layout}).  The MACS-1149 \HAWKI\ observations were obtained in Service Mode at the ESO/\textit{VLT} in between 2013 Mar 21 and 2014 Jun 9.  The 20$\times$15\,s~=~300\,s exposures were taken at nine dithered positions offset within a $30\arcsec$ box; the on-source exposure time of the final MACS-1149 \HAWKI\ mosaic is 5.3 hours.

\subsection{Image processing}
\label{s:processing}

The \HAWKI\ and MOSFIRE observations were reduced with a pipeline that has been developed for previous surveys with the NEWFIRM \citep[NMBS;][]{whitaker:nmbs} and FOURSTAR (ZFOURGE; \citealt{spitler:12}; Straatman et~al, submitted) infrared imaging instruments.  Treating each detector individually, the pipeline is easily modified for the different instrument configurations of the four \HAWKI\ and single MOSFIRE chips.  The primary task of the pipeline is removing the bright, time-variable sky background from the individual exposures, which is often some $10^4$ times brighter than the distant galaxies of interest in the field.  With such a bright background, we first determine an empirical ``sky flat'' that is a median of all of the \textit{science} exposures in a \HAWKI\ OB or MOSFIRE group, after rejecting the brightest 12 exposures at each pixel position to remove the contribution of bright objects.  We find these empirical flats to be preferable to external twilight or dome flats given the difficulty of obtaining a truly flat illumination pattern over such large detector fields-of-view.  

After dividing by the flat, the background of each exposure is determined in a first pass from the simple median of the four exposures that came both immediately before and after it.  The first-pass background-subtracted exposures are combined into a mosaic, and objects are identified as pixels with values greater than five times the robust standard deviation \citep{beers:90} of the combined image.  A buffer with radius 3 pixels is grown around each ``object'' pixel that satisfies this 5$\sigma$ criterion.  The final refined background of each exposure is determined in a second ``mask pass'' from a median again from the four exposures before and after it but now masking all pixels that contain flux from the detected objects.

As the archival \HAWKI\ images of the MACS-1149 field were obtained with longer individual exposures and with fewer exposures per sequence, the mask-pass technique described above did not produce satisfactory results.  In this case, for each raw exposure we divide by the empirical sky flat and then subtract a third-order polynomial fit to the background.  

\begin{figure}[!t]
\epsscale{1.17}
\plotone{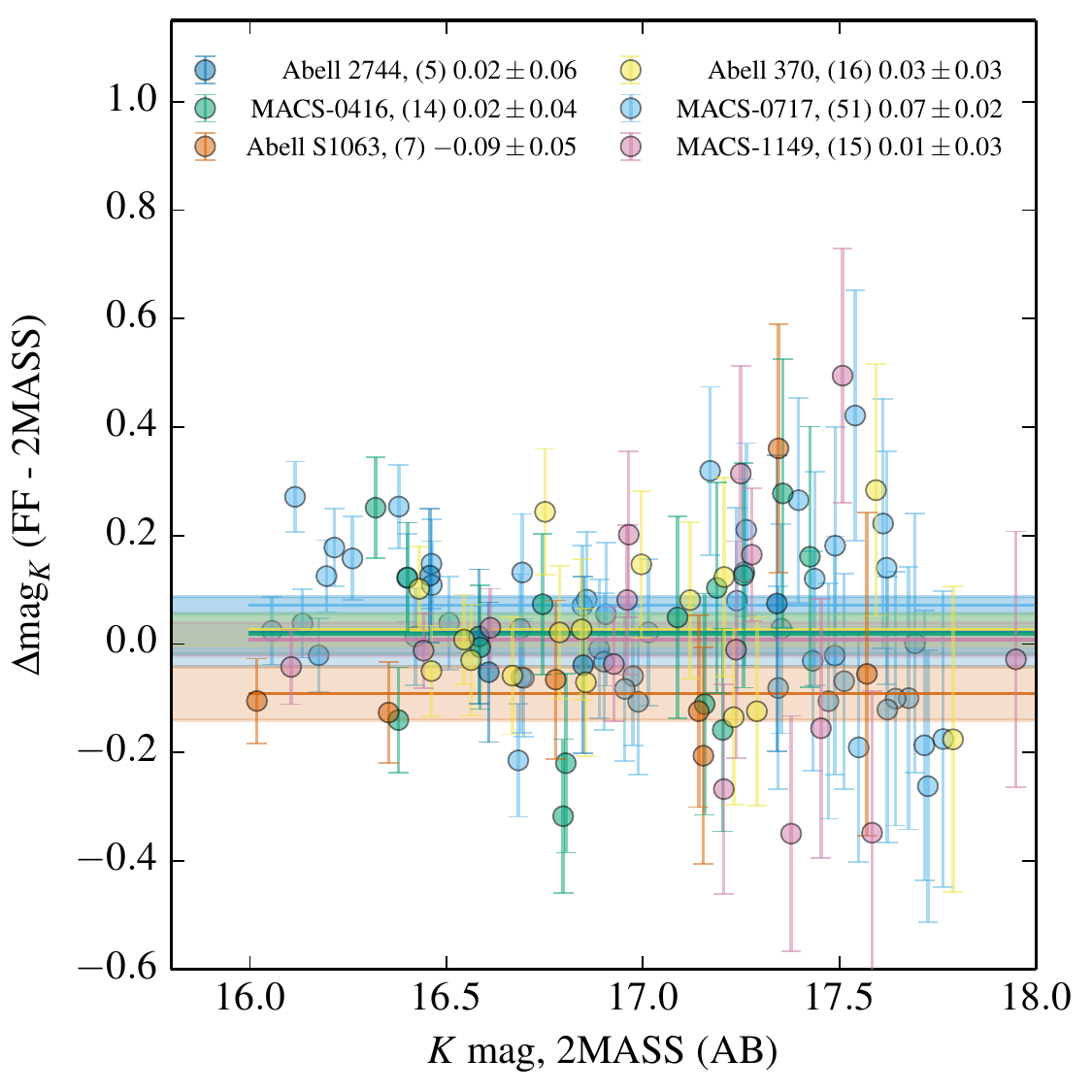} 
\caption{Comparison of bright-star photometry to the 2MASS public catalog.  The photometric calibration of the $K_s$-band images were determined from standard star observations taken concurrently with the science exposures.  The uncertainties shown are taken from the 2MASS catalog; these stars are measured at very high S/N in the deep $K_s$-band mosaics.  The solid lines and shaded regions show the weighted average and standard deviation of the photometric offsets for each field, with the quantitative values indicated in the labels at the top of the figure.  \label{fig:2mass}}  
\end{figure}

\subsection{Photometric calibration}
\label{s:calib}

For the \HAWKI\ observations, a single 90-minute Service Mode OB was obtained in each of the fields requiring photometric transparency conditions, and these OBs were followed immediately by an observation of a photometric standard star at a similar airmass with single exposures on each of the four chips.  The standard star exposures were processed with empirical sky flats derived from the science exposures as described above, and the observed fluxes of the standard stars yield absolute photometric zeropoints for each chip.  Correction factors were then computed to scale the OBs obtained at varying transparency levels to the calibrated photometric OB; the additional service mode OBs were obtained under generally satisfactory (i.e., ``clear'') weather conditions and the scale factors typically differ from unity by only a few percent.  The photometric calibration of the MOSFIRE observations was determined from standard-star observations taken the same night as the science exposures, again with the standard exposures reduced in the same way as the science exposures.

The 2-Micron All Sky Survey \citep[2MASS;][]{2mass} provides an additional check on the photometric calibration, though the comparison is limited due to relatively little brightness overlap between the faint end of reliable 2MASS photometry and the bright end where stars are in the linear regime of the detectors on the 8--10\,m telescopes.  A comparison of the observed photometry in the four survey fields to the 2MASS catalog magnitudes is shown in Fig.~\ref{fig:2mass}.  The stellar photometry on the deep $K_s$-band mosaics is measured within $1\arcsec$ apertures corrected to infinity with the curves of growth described below.  The error-bars on the points in Fig.~\ref{fig:2mass} come from the 2MASS catalog; the photometric uncertainties are negligible for such bright stars in the deep images ($\mathrm{S/N}>100$).  The 2MASS comparison suggests a systematic zeropoint offset of $-0.09$~mag for the Abell S1063 field.  However, the exposure time for that field was similar to those of the additional deep \HAWKI\ fields and the derived depths of \textit{all} deep \HAWKI\ fields are nearly identical (Tab.~\ref{tab:summary}).  Therefore, we do not apply this offset to the S1063 zeropoint since if real it should be reflected in the derived depth of that field as compared to the others.  Given the overall good agreement with the 2MASS photometry, we estimate that the systematic uncertainty of the absolute photometric calibration is $\sigma_\mathrm{sys}\leq0.05$~mag.  

Similar to \cite{hugs}, we do not apply any correction for non-linearity effects of the \HAWKI\ or MOSFIRE detectors.  For the case of \HAWKI\, the user manual suggests that the detector is linear at the 1\% level up to 30~000 detector counts (ADUs).  For a 2MASS star in the Abell~2744 field with $K=17.3$, the brightest pixel reaches $\sim$31~000~ADUs including the bright sky background.  Therefore, there could be a linearity correction of up to a few percent between the bright 2MASS stars used for the photometric comparison and galaxies in the field some 8--9 magnitudes fainter, though the stars in Fig.~\ref{fig:2mass} do not show any significant slope in the 2MASS magnitude residuals across $\sim$1.5~mag of dynamic range.  For the fainter galaxies, empirical ``zeropoint corrections'' are often computed as part the photometric redshift analysis \citep[e.g., ][]{skelton:14}, and these would compensate for relatively small linearity effects.  

\subsection{Astrometric alignment}
\label{s:astrometry}

In order to simplify measurements from the $K_s$-band images in consort with space-based \textit{HST} and \textit{Spitzer} mosaics, the astrometry of the ground-based mosaics must be refined.  Reference absolute astrometric catalogs were generated from \textit{HST} images when available (i.e., the chips overlapping the cluster and parallel \textit{HST} fields) and public \textit{Subaru} Suprime-Cam $r_c$-band images otherwise\footnote{\url{http://archive.stsci.edu/pub/hlsp/clash/}}.  Next, object catalogs are generated with the SEXtractor software \citep{bertin:96} for each individual background-subtracted exposure and transformations to the reference frame (shift, rotation and scale) are computed with the \texttt{stsci.stimage.xyxymatch} and \texttt{scikit-image.transform}\footnote{\url{http://scikit-image.org}} Python software tools.  For the \HAWKI\ exposures, we fit a third-order polynomial to the geometric distortion model determined by \cite{libralato:14} and specify the polynomial terms as ``Simple Imaging Polynomial'' (SIP) coefficients \citep{fits:sip} in the individual exposure FITS files.  The SIP FITS header generally applicable for \HAWKI\ is provided in Appendix~\ref{appendix:sip}.  We note that the \HAWKI\ distortion is generally small, reaching $\sim$2 pixels ($0\farcs2$) at the image corners, but that the distortion corrections are necessary, in particular, given the excellent overall image quality of the observations.

\begin{figure}[!t]
\epsscale{1.17}
\plotone{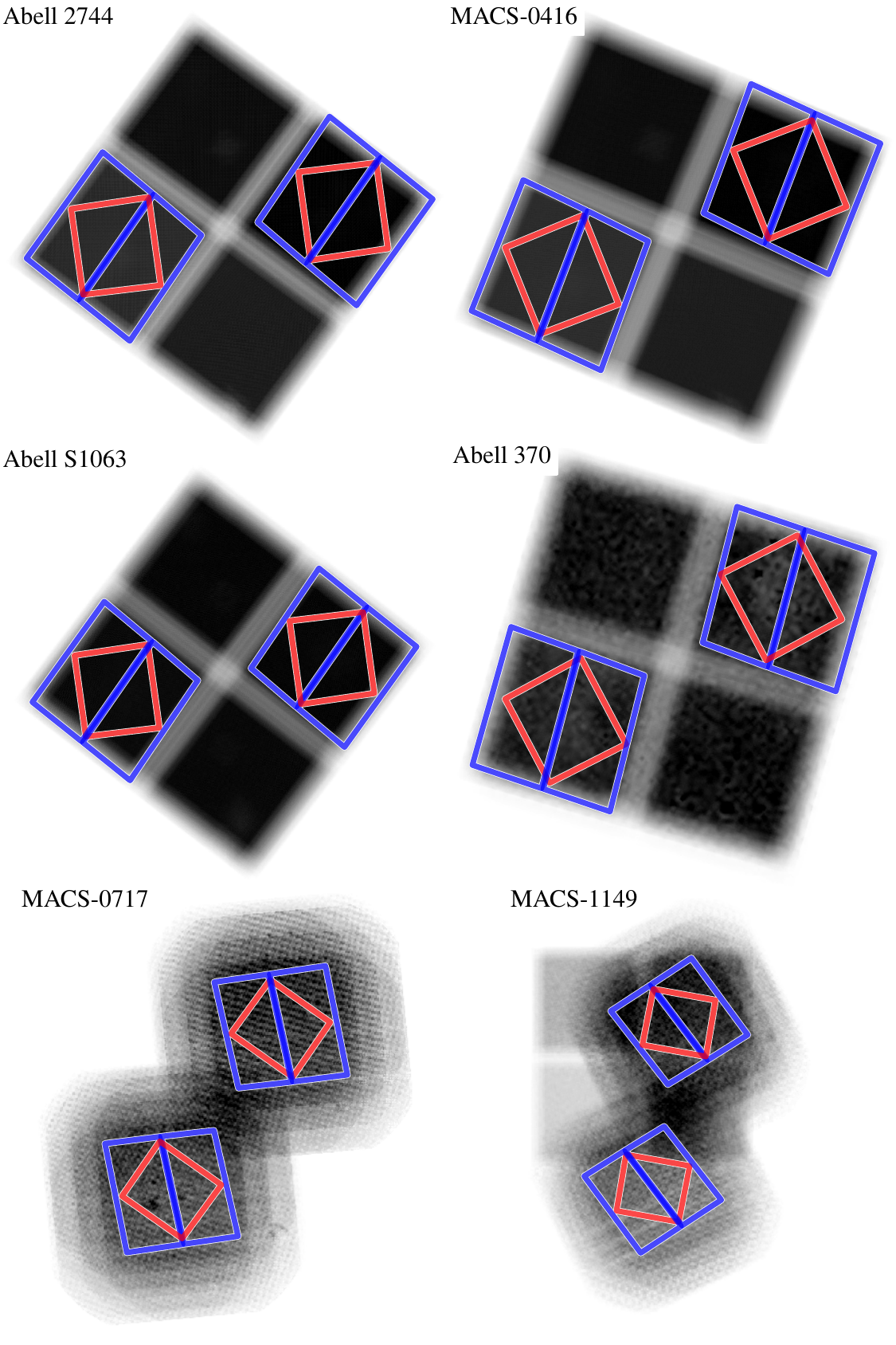} 
\caption{Inverse variance maps of the $K_s$-band mosaics.  The orientation of the fields is the same as that shown in Fig.~\ref{fig:layout}.  The four \HAWKI\ chips and two MOSFIRE pointings are clearly visible for the first four and last two fields, respectively.
\label{fig:ivar_mosaics}}  
\end{figure}

\section{Drizzled $K_s$-band Mosaics}
\label{s:mosaic}

\begin{figure*}[!t]
\epsscale{1.17}
\plotone{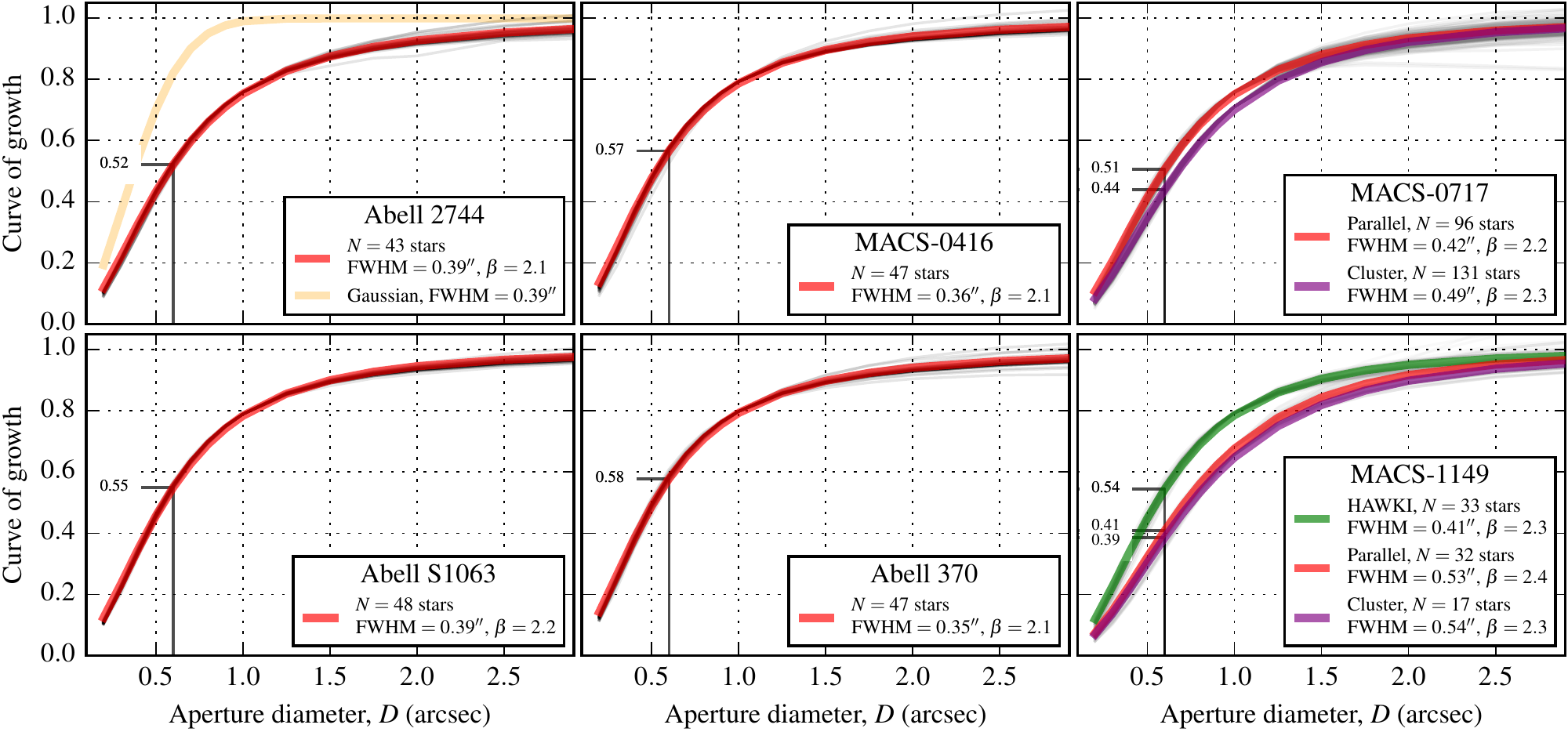} 
\caption{Curves of growth of stellar profiles in the Frontier Fields $K_s$-band mosaics.  The \HAWKI\ fields include point sources across the full mosaics that cover both \textit{HST} cluster and parallel fields, while separate curves are given for the cluster and parallel areas of the northern MOSFIRE fields, which required two independent pointings of that instrument (Fig.~\ref{fig:layout}).  The inverse aperture corrections for point sources in $D=0\farcs6$ apertures are indicated in each panel with the small labels.  The profiles of stars are well-fit by Moffat profiles with $\mathrm{FWHM}\sim0\farcs4$ in most cases and $\beta\sim2.1$.  These extended Moffat profiles appear to be characteristic of deep $K_s$-band images and they have substantially more flux at large radii ($r>\mathrm{FWHM}/2$) than Gaussian profiles with the same FWHM, as shown in by the orange curve in the upper-left panel). \label{fig:cog}}  
\end{figure*}

A crucial innovation of the image processing of the present analysis compared to previous works is that we combine all of the individual background-subtracted raw images (e.g., 7040 files for 1760 exposures $\times$ 4 detectors for the \mbox{Abell 2744} field) into the final output mosaic with the ``drizzle'' algorithm (\citealp{fruchter:02}, as implemented in the DrizzlePac/AstroDrizzle package; \citealp{drizzlepac}).  The benefits of the drizzle algorithm  preserving robust noise properties of the output mosaics will be discussed in detail below in \S\ref{s:depth}.  Here we simply indicate that the drizzle implementation provides additional benefits in 1) trivial application of relative weights of the input images in creating the final mosaics, 2) providing infrastructure to apply the astrometric alignment and geometric distortion stored in the FITS headers, and 3) allowing the definition of an arbitrary output pixel grid without any additional image resampling.

Final mosaics of each field with $0\farcs1$ pixels are drizzled from the input exposures with weights \citep[following][]{whitaker:nmbs}
\begin{equation}
w^{-1} = F^2\cdot B \cdot \mathrm{FWHM}^\alpha / \sqrt{1-e^2}, 
\label{eq:weight}
\end{equation}
where $F$ is the factor to scale the exposure to the photometric system ($F\equiv1$ for the average of exposures in the OB taken under photometric conditions, $F\gtrsim1$ otherwise), $B$ is the measured background, FWHM is the full width at half maximum of stars identified in the field, and $e$ is the ellipticity of the stellar point spread function (PSF).  The parameter $\alpha$ allows for optimizing the image quality of the final mosaic.  Increasing $\alpha$ puts larger weight on the exposures with the best image-quality while effectively ignoring data with poorer seeing; we adopt $\alpha=2$ for a compromise between the image quality and effective exposure time of the final mosaic.  

Finally, we subtract a cell-based background from the final mosaics using an algorithm based on that used by SExtractor and SWarp but that provides more aggressive masking of flux in the outer isophotes of bright galaxies. The mosaics are shown in Fig.~\ref{fig:layout}, along with the position of the deep Frontier Fields \textit{HST} imaging fields.  We also compute the robust NMAD scatter \citep{brammer:08} of empty sky pixels within the same cells to empirically calibrate the inverse variance maps that are shown in Fig.~\ref{fig:ivar_mosaics} (see also \S\ref{s:depth}).  The final science and inverse variance mosaics of all six fields are provided as ESO Phase 3 data products\footnote{\url{http://www.eso.org/sci/observing/phase3/news.html\#kiff}}.  The released images are all scaled to a common zeropoint of 26.0 (AB mag), which gives pixel values of order unity faint galaxies near the detection threshold.  In the sections below we describe the characteristics of the mosaics in more detail.

\subsection{Image quality}
\label{s:fwhm}

Fig.~\ref{fig:cog} shows the curves of growth for stars identified by the tight relationship between their brightness and half-light radii \citep[see, e.g., Fig.~13 of][]{skelton:14}.  The image quality of the $K_s$-band mosaics is excellent with stellar FWHM$\lesssim0\farcs4$ for the deep \HAWKI\ fields, thanks in large part to the execution of the \HAWKI\ observations in service mode ensuring optimal and uniform image quality across the many hours of integration on the survey fields.  The classical scheduling of the MOSFIRE observations does not allow such control over the seeing conditions.  The result is that the image quality of the MACS-0717 and MACS-1149 fields is somewhat degraded in comparison, at $0\farcs42$--$0\farcs54$, and the image quality differs measurably between the two MOSFIRE pointings in the MACS-0717 field.  Despite the compact cores of the stellar PSFs, the curves of growth shown in Fig.~\ref{fig:cog} deviate from Gaussian profiles with significantly higher flux in the outer wings, with a shape more consistent with a Moffat profile \citep{trujillo:01}  with $\beta\sim2$.  We caution that these extended profiles will yield significantly shallower final image depths than would be predicted with Exposure Time Calculators that may assume Gaussian profiles.  Nevertheless, the deep $K_s$ mosaics have excellent image quality that is much closer to the resolution of the \textit{HST} IR imaging and the typical apparent size of distant galaxies \citep[median $r_e\sim0\farcs1$--$0\farcs2$ at $z>4$;][]{shibuya:15} than the redder \textit{Spitzer} IRAC bands (cf. FWHM$\sim$$1\farcs7$--$2^{\prime\prime}$).

The image quality obtained here with HAWK-I demonstrates the excellent natural seeing conditions of the \textit{VLT} site at Cerro Paranal.  We note here that \HAWKI\ will soon be upgraded with a ground layer adaptive optics module \citep[GRAAL;][]{paufique:10}, that will improve the image quality by factors of 1.5--2 over the natural seeing over the wide \HAWKI\ field of view.  \cite{schirmer:15} recently presented deep AO-corrected $K_s$-band imaging of the MACS-0416 cluster field obtained with the Gemini South Adaptive Optics Imager \citep[GSAOI;][]{carrasco:12}.  Thanks to the availability of a bright guide star near the center of the cluster field, GSAOI provided exquisite image quality (FWHM$\sim0\farcs09$) better than even that of \textit{HST} WFC3/IR.  Though somewhat shallower ($K_s\sim$25.6, 5$\sigma$) and covering a smaller field of view ($1\farcm7\times1\farcm8$, Fig.~\ref{fig:layout}) than the \HAWKI\ mosaics, the GSAOI images provide an auspicious demonstration of the power of wide-field AO-corrected imaging for deep extragalactic science.

\subsection{Noise properties \& Depth}
\label{s:depth}

\begin{figure}[!t]
\epsscale{1.17}
\plotone{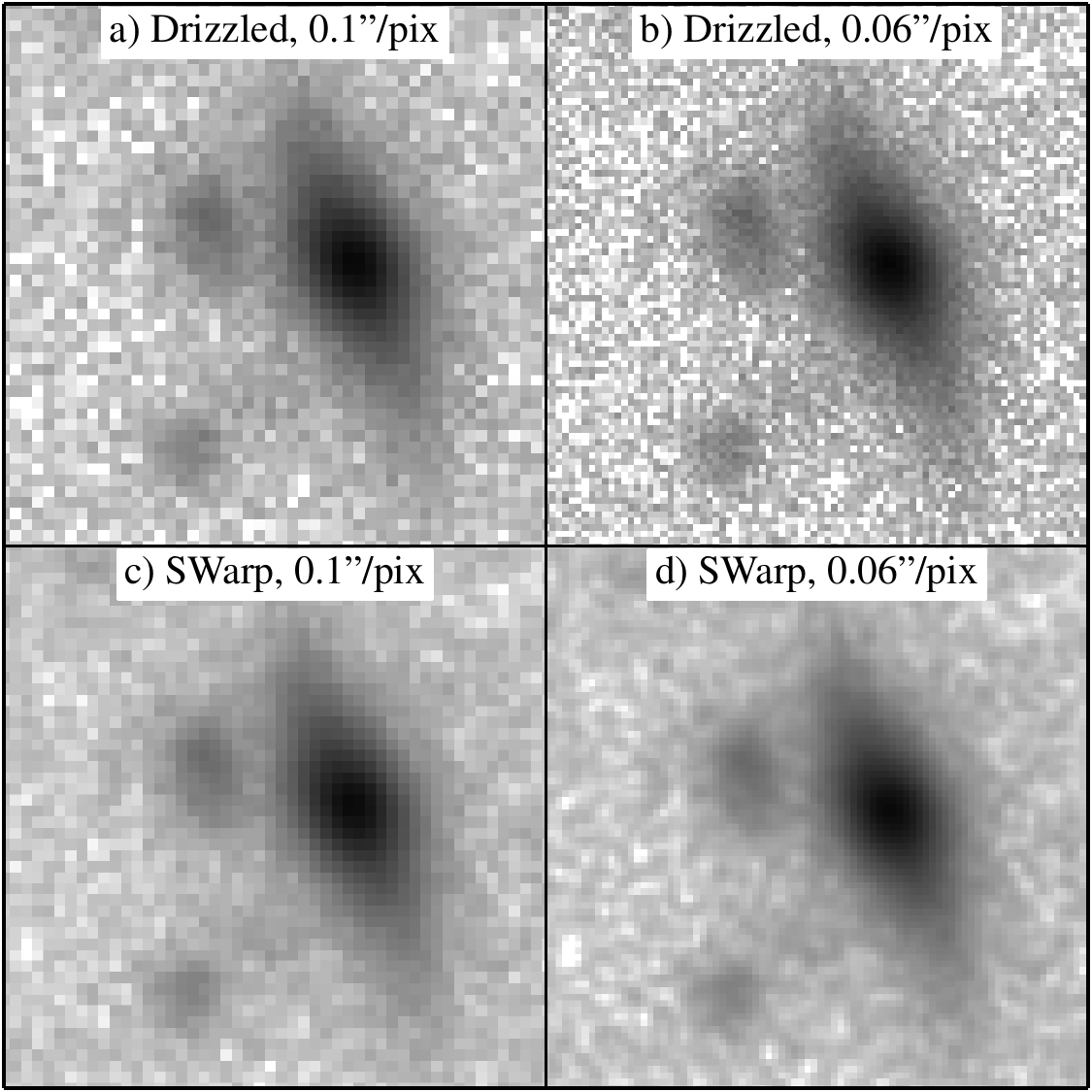} 
\caption{Comparison of small regions of the final MACS 0416 mosaics with two output pixel sizes for images combined with the drizzle algorithm (top panels) and with the SWarp software (bottom panels).  With drizzle we can decrease the size of the input pixels before dropping them into the output grid, which significantly reduces the correlations between neighboring pixels.  Combining images with simple shift-and-add resampling such as with SWarp results in an effective smoothing of the pixel-to-pixel noise in the final combined image. 
\label{fig:drizzle_cutout}}  
\end{figure}

\begin{figure}[!t]
\epsscale{1.17}
\plotone{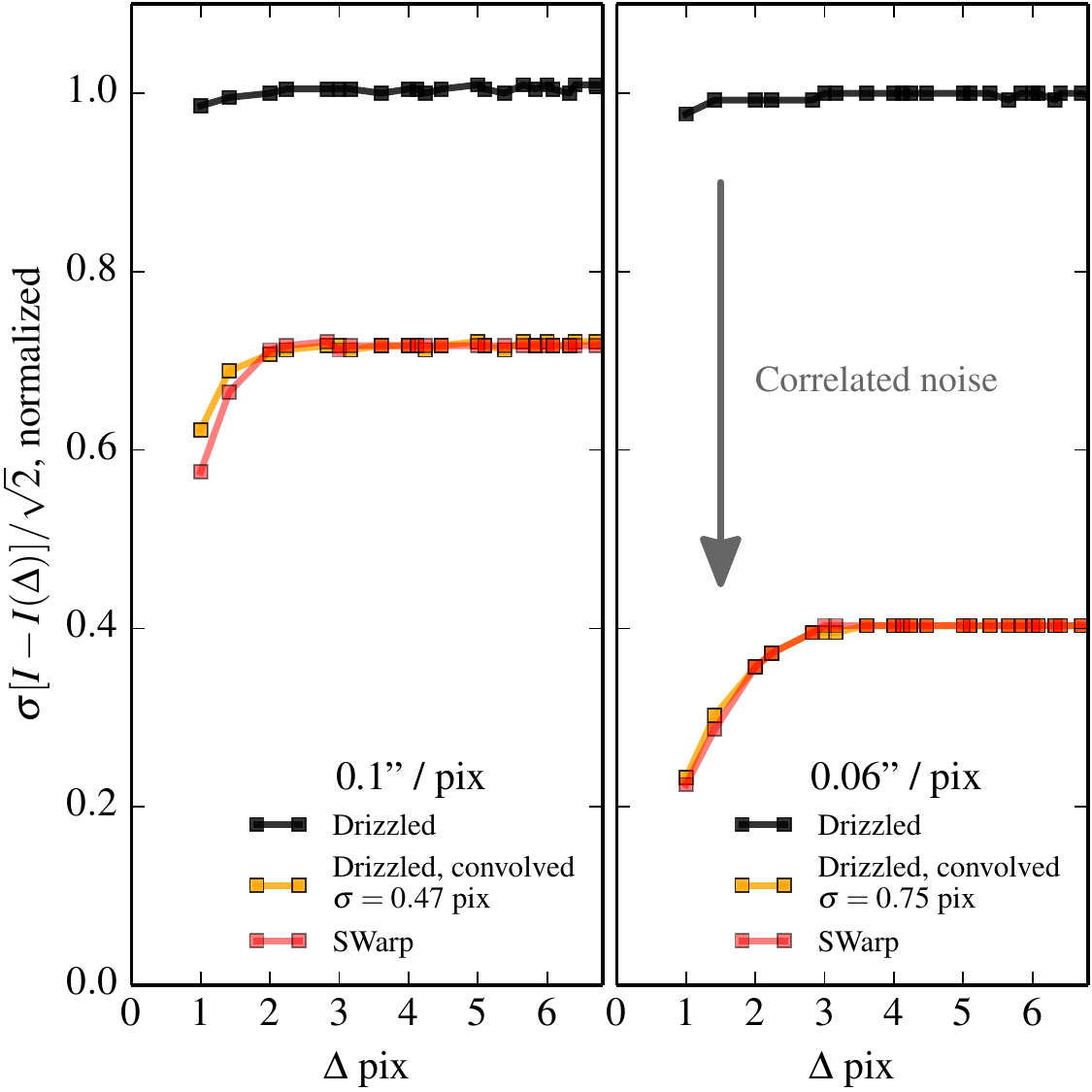} 
\caption{The standard deviation of the pairwise pixel differences, sorted by pixel separation, provides a quantitative measure of the correlations between pixels.  The black curves determined from the drizzled images show very little variation with pixel separation, suggesting minimal correlations between adjacent pixels, as expected.  In contrast, the statistics of the SWarped images show a depression at small separations indicative of correlated pixels.  The magnitude and shape of the depression is reproduced exactly by the orange curves, which show the pairwise differences for the drizzled images convolved with small Gaussian kernels.  The effect is larger in the right panel, where the ratio of the input and output pixel sizes is larger by almost a factor of two.
\label{fig:corr_noise}}  
\end{figure}

Along with practical benefits mentioned above, the most dramatic benefit of using the drizzle algorithm comes from the fact that the raw detector pixels are only resampled once in the process of generating the output mosaic.  Furthermore, with a large number of dithered raw images we can use drizzle to shrink the raw input pixels by a factor of ten before dropping them into the output grid,\footnote{The pixels are shrunk by adopting pixfrac=0.1 as defined by \cite{fruchter:02}.} and the result is dramatically reduced correlations between neighboring pixels, particularly for output pixel sizes that are somewhat smaller than the native \HAWKI\ or MOSFIRE detector pixels.

A comparison of final combined images created with output $0\farcs1$ and $0\farcs06$ pixel grids using the drizzle implementation to versions created with the SWarp software \citep{swarp} is shown in Fig.~\ref{fig:drizzle_cutout}.  In the later case, the raw pixels are resampled twice before making the final mosaic, once combining the raw exposures in each OB and a second time combining the separate OBs into the final stack with an arbitrary pixel grid.  One could avoid the second resampling and combine all of the raw exposures at once, though SWarp is still limited to dropping the original pixels into the output mosaic preserving the input native pixel grid and therefore a single input pixel maps onto multiple (perhaps many) output pixels.

The fact that the drizzle-combined images in Fig.~\ref{fig:drizzle_cutout} (panels \textit{a} and \textit{b}) appear to the eye to be noisier than their SWarped counterparts (panels \textit{c} and \textit{d}) is simply the result of a smoothing of the pixel noise in the latter case rather than reflecting true underlying differences in their pixel-to-pixel variance.  \cite{casertano:00} discuss how the noise properties of correlated output pixels are affected by the relative sizes of the input and output pixel grids and the adopted drizzle parameters.  With the drizzle algorithm and a small pixfrac, a given input pixel will map to only a single output pixel and therefore correlations between the output pixels should be minimal\footnote{Inter-pixel capacitance and other forms of crosstalk intrinsic to the IR detectors place a lower limit on the correlations between adjacent pixels \citep{finger:05, hilbert:ipc}.}.  

We explore these correlations between the output pixels in Fig.~\ref{fig:corr_noise}.  For the black and red curves in each panel, we measure value differences between random pairs of pixels sorted by the separation between the pixels for the drizzled and SWarped images, respectively.  For perfectly uncorrelated pixels, these curves will be flat and will reflect the intrinsic noise of the image pixels.  In the presence of correlations between adjacent pixels, however, the curves will show a depression at small separations.  As expected, the SWarp-combined images show exactly such a depression for pairs of pixels 1--2 pixels apart.  The yellow curves in Fig.~\ref{fig:corr_noise} show the same pairwise differences for the drizzled images now convolved by small Gaussian kernels as indicated, which agree very well with the curves for the images with correlated pixels.  

In terms of understanding the noise properties of the images, a key point here is not only that depression at small separations indicates the presence of correlations between adjacent pixels but also the fact that correlations cause the apparent overall r.m.s. to be decreased at all separations.  That the pixel variances cannot be used directly in the presence of such correlations \citep{casertano:00} is the primary reason why significant effort has been devoted to placing random ``empty apertures'' across images in order to characterize their noise properties \citep[e.g.,][]{labbe:03,forster:06,quadri:musyc, whitaker:nmbs, skelton:14}.  Now by eliminating the pixel correlations with the drizzle image combination technique (with a small pixfrac) we have shown that the pixel statistics are robust and it is now trivial to compute the expected variance within an arbitrary photometric aperture, for example:
\begin{equation}
\sigma_{\mathrm{aper}, D}^2 = \sigma_\mathrm{pix}^2\cdot \pi/4 \cdot D^2,
\label{eq:aper_var}
\end{equation}
for circular apertures with diameter, $D$, in pixels, and where $\sigma_\mathrm{pix}^2$ is the per-pixel variance determined from an analysis such as that in Fig.~\ref{fig:corr_noise}.

\begin{figure}[!t]
\epsscale{1.17}
\plotone{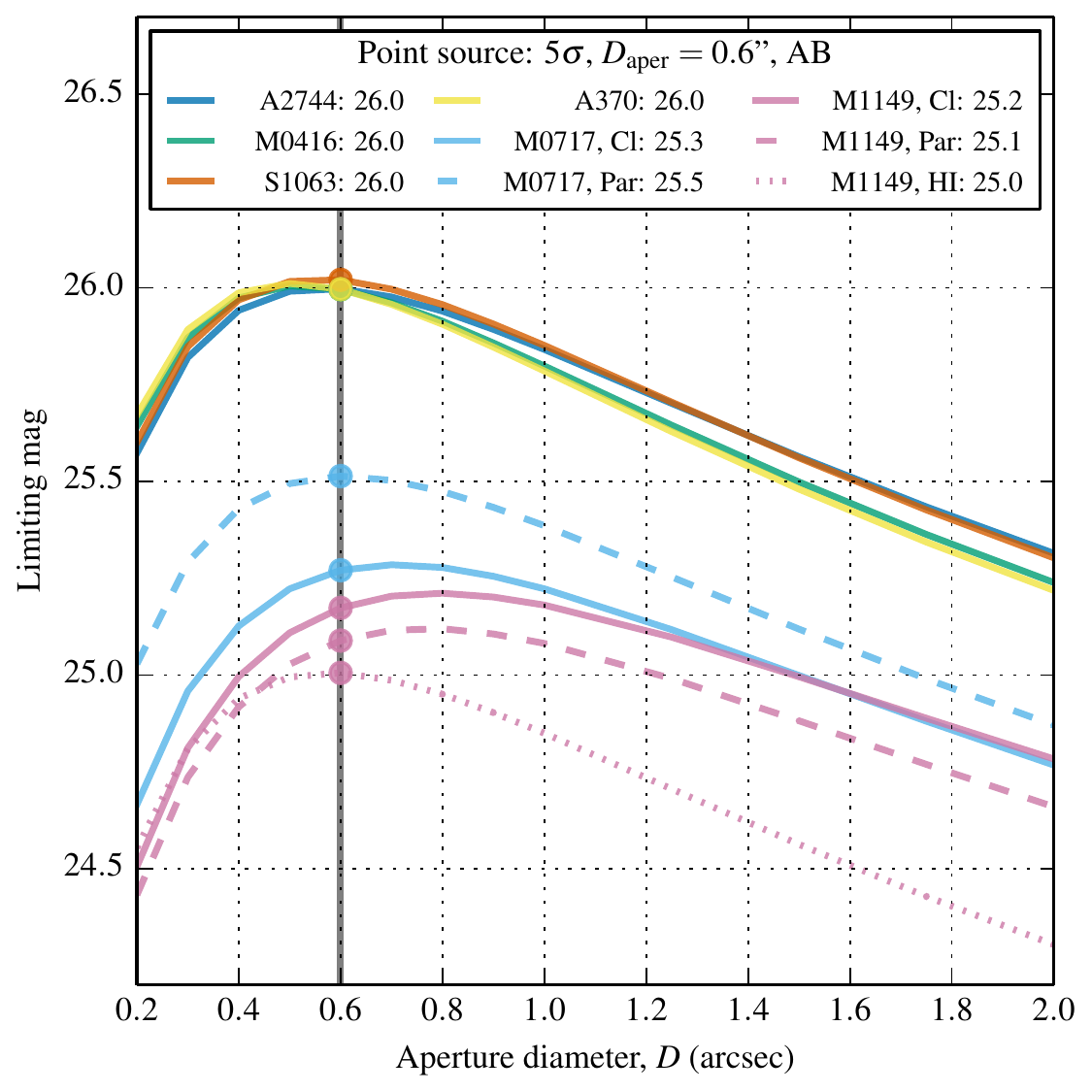} 
\caption{The product of the inverse curves-of-growth (Fig.~\ref{fig:cog}) and the predicted variance within a photometric aperture (Eq.~\ref{eq:aper_var}) gives the point-source sensitivity as a function of aperture size.  The S/N is maximized with an aperture somewhat larger than the FWHM; depths evaluated at $D=0\farcs6$ near the maxima are indicated.)
\label{fig:depth}}  
\end{figure}

\begin{figure}[!t]
\epsscale{1.17}
\plotone{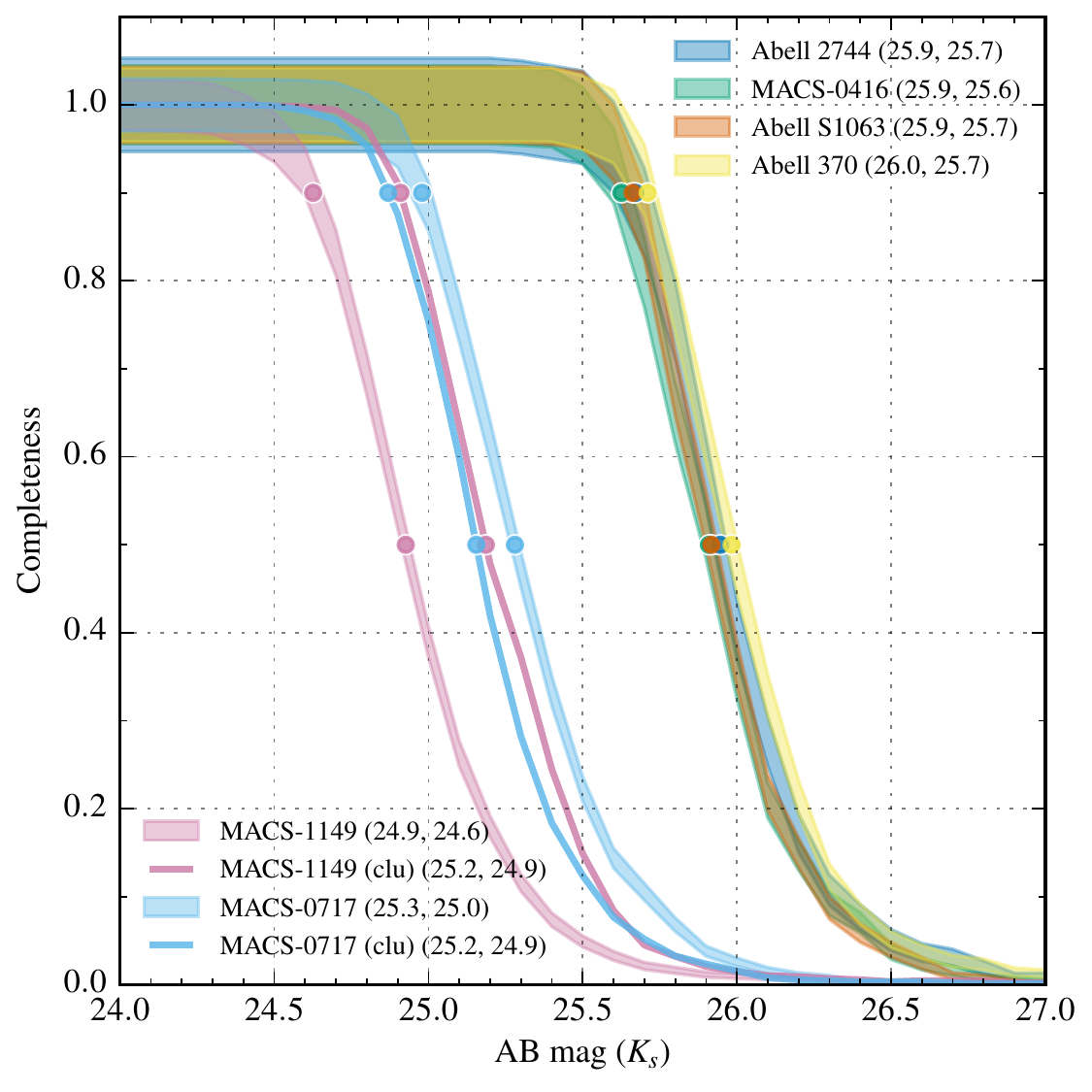} 
\caption{$K_s$-band detection completeness in the six Frontier Fields survey fields. The four deep \HAWKI\ fields are remarkably uniform, with 50\% detection completeness at $K_s=25.9$--26.0 AB.  The shallower northern MACS-1149 and MACS-0717 fields are 50\% complete at $K_s=24.9$ and 25.3, respectively.  Separate curves are shown for the cluster and parallel areas in the MACS-0717 and MACS-1149 fields, which required two separate pointings with the MOSFIRE instrument;  the completeness curves in the cluster areas of the four \HAWKI\ fields differ from the curves shown by $\lesssim0.1$ mag and are omitted for clarity.
\label{fig:completeness}}  
\end{figure}

The product of Equation~\ref{eq:aper_var} and the inverse of the curves of growth shown in Fig.~\ref{fig:cog} (i.e., aperture corrections) defines an optimal photometric aperture size where this product is minimized \citep{whitaker:nmbs}.  For our $K_s$-band mosaics, this optimal aperture has $D\sim0\farcs6$, just larger than the FWHM of point-source profiles.  The $5\,\sigma$ limiting magnitudes (AB) within $0\farcs6$ diameter apertures are indicated in Fig.~\ref{fig:depth} and listed in Table~\ref{tab:summary}.  Reaching 26th magnitude (AB), the \HAWKI\ images presented here are among the deepest $K_s$-band images ever obtained and will provide an important complement to the deep \textit{HST} and \textit{Spitzer} imaging of the Frontier Fields.  

We conclude here with a brief comment on the ``HUGS'' ultra-deep \HAWKI\ images of the CANDELS GOODS-S and UDS survey fields.  \cite{hugs} report a final depth of 26.5 AB for point sources within $D=0\farcs4$ apertures for their deepest field, ``GOODS-D1''.  This field has an exposure time of 31.5 hours, similar to the deep \HAWKI\ fields summarized in Table~\ref{tab:summary}.  \cite{hugs} assume ideal noise properties that likely suffer some residual correlation between adjacent pixels such as that indicated by the red curves in Fig.~\ref{fig:corr_noise}, where we found that the drizzled rms was some 40\% higher than that measured in the presence of the pixel correlations for $0\farcs1$ pixel mosaics.  This effect ($\sim$0.4 mag) along with the relative exposure time difference ($\sim$0.1 mag) can account for much of the difference between the reported depths of the deepest HUGS pointings and the \textit{Hubble} Frontier Fields \HAWKI\ images described here.  The comparison must be made because any additional half magnitude increase depth is exponentially more expensive to obtain!

\subsection{Source detection \& completeness}
\label{s:completeness}

We compute source detection completeness curves as a function of source brightness following the techniques described by \cite{whitaker:nmbs} and \cite{muzzin:13}.  Briefly, we insert artificial sources in blank regions of the portions of the images covering the \textit{HST} ``parallel'' fields and compute the fraction of sources recovered with SExtractor as a function of the source magnitude.  The resulting completeness curves are shown in Fig.~\ref{fig:completeness}.   As in the sensitivity analysis above, we find that the detection completeness curves are nearly identical for the deep \HAWKI\ fields, with 50\% (90\%) source completeness at $K_s\sim25.9$ (25.7) AB.  The completeness thresholds for the shallower northern MACS-1149 and MACS-0717 fields are $\sim$0.75 mag brighter.

We perform this completeness analysis to provide a general characterization of the $K_s$-band image mosaics and to provide a point of comparison for earlier surveys, such as UltraVISTA \citep{muzzin:13}.  However, in the case of the Frontier Fields cluster and parallel fields with much deeper photometry available from \textit{HST} (c.f. 28.7 AB in $H_{160}$), the $K_s$ band images can be more effectively exploited for most applications by detecting objects in the deeper \textit{HST} images and performing forced $K_s$-band photometry at the positions of the \textit{HST} sources.  \cite{merlin:16} present multiwavelength catalogs of the Abell~2744 and MACS-0416 fields constructed in this way, including incorporating an early version of the \HAWKI\ $K_s$-band mosaics.  The generation of these catalogs is beyond the scope of the present work and will be presented in further detail by Shipley et al. (2016; in preparation).

\begin{figure}[!t]
\epsscale{1.17}
\plotone{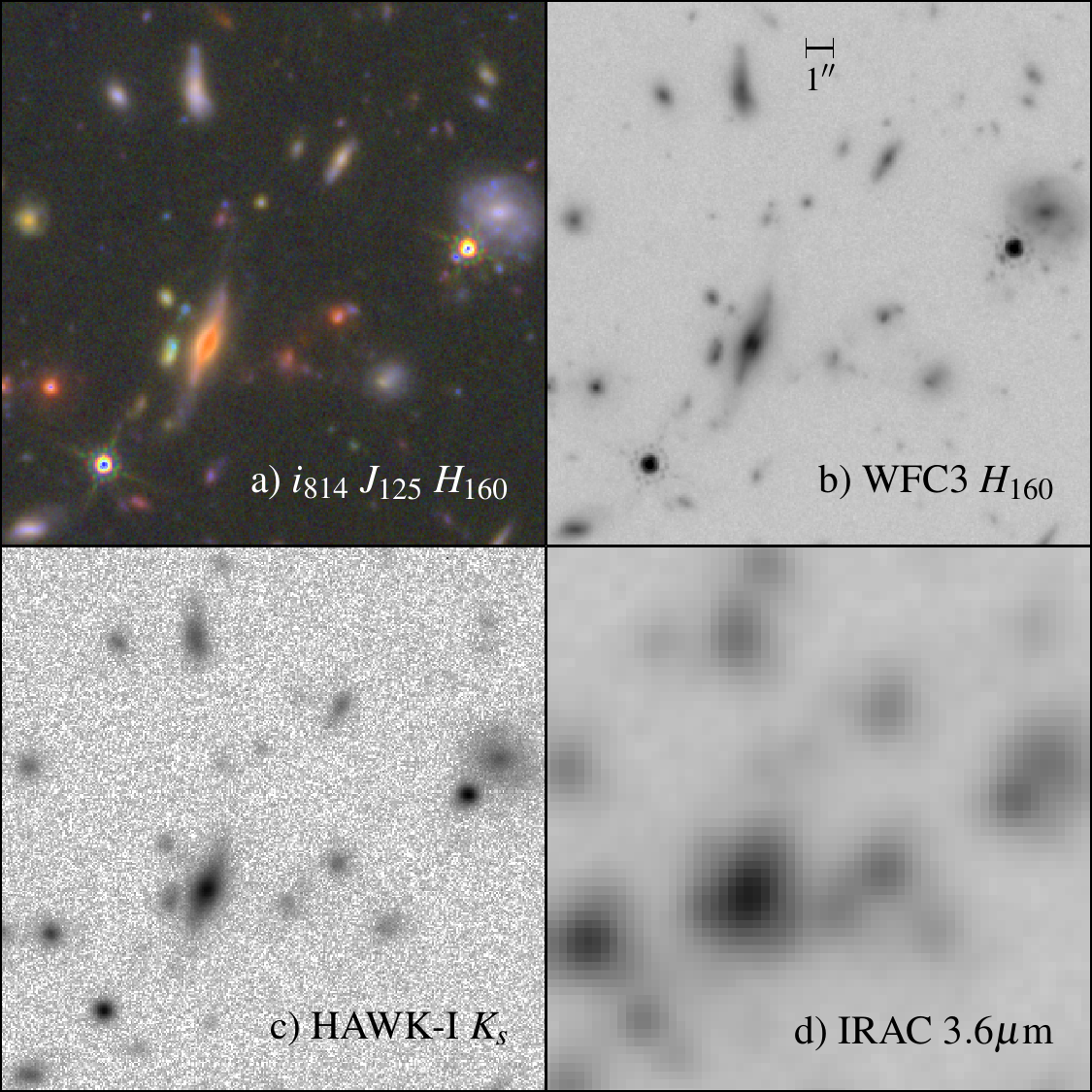} 
\caption{Image cutouts of the MACS 0416 parallel field.  The \textit{RGB} and monochrome panels come from the sources as indicated; the cutouts are all $20\arcsec$ on a side.
\label{fig:cutout}}  
\end{figure}

\begin{figure*}
\epsscale{1.17}
\plotone{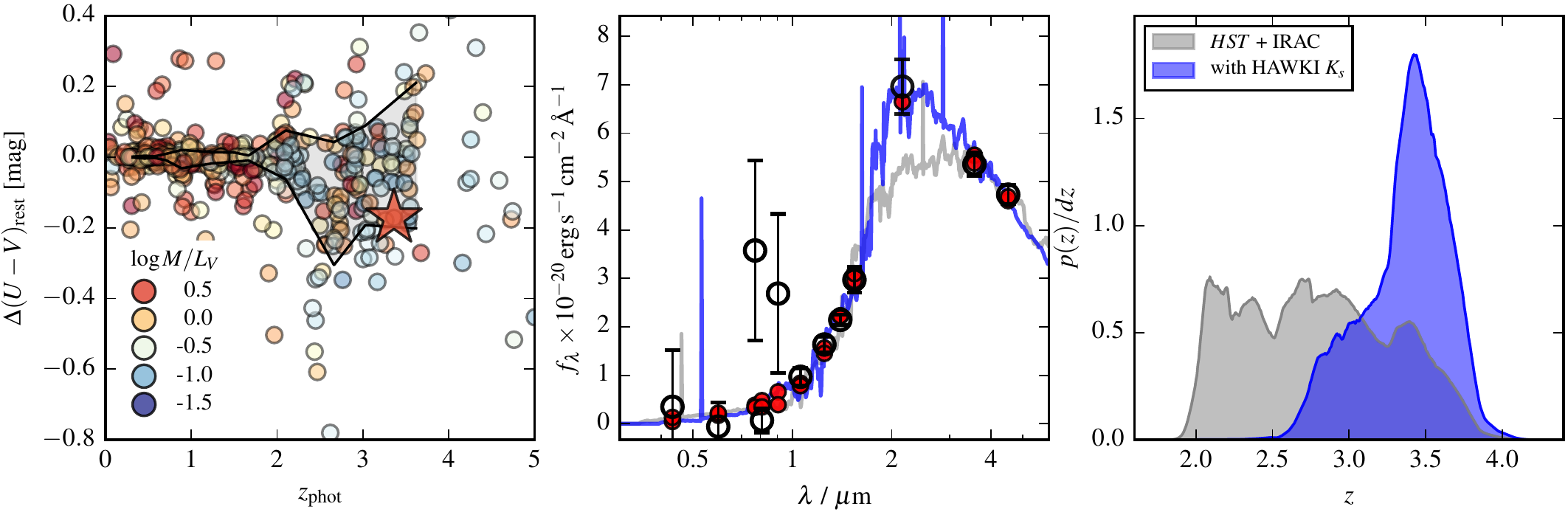} 
\caption{\textit{Left:} Difference in the derived rest-frame $U-V$ colors as a function of photometric redshift for galaxies in the MACS-0416 parallel field, before and after including $K_s$ photometry along with the \textit{HST} and \textit{Spitzer} measurements.  The scatter in the derived $U-V$ colors (solid black lines) is large at $z>2.5$ ($\sim$0.2 mag) where the Balmer/4000~\AA\ break is in the gap between \textit{HST} $H_{160}$ and the \textit{Spitzer} IRAC 3.6~\mum\ bands, which is then well constrained by including $K_s$.  This large scatter translates directly into large scatter on other quantities of interest, such as the stellar mass to light ratio and the age of the stellar populations.  The center and right panels show a dramatic example of a galaxy with a strong break between the \textit{HST} and IRAC bands.  Its photometric redshift and $\Delta(U-V)$ color are indicated with the large red star in the left panel.  The SED fits (center panel) and derived photometric redshift probability densities (right panel) before and after including the $K_s$-band photometry are shown with gray and blue curves, respectively.  The addition of the $K_s$ band pinpoints the location of the break and shrinks the photometric redshift uncertainties by a factor of two and suggesting $z_\mathrm{phot} > 3$ (right panel). \label{fig:delta_UV}}  
\end{figure*}



\section{Discussion and Conclusions}
\label{s:discussion}

We have presented the observations and reduction for a collection of extremely deep $K_s$-band images covering the lensing clusters observed as part of the \textit{Hubble} and \textit{Spitzer} Frontier Fields program (Lotz et al., in prep.; Capak et al., in prep.).  Service mode scheduling of $>$25~hour \HAWKI\ integrations on the southern hemisphere clusters visible from the ESO/\textit{VLT} result in remarkably uniform, high-quality mosaics across the separate survey fields, with superb image quality ($0\farcs4$~FWHM) and photometric depths reaching AB=26.0 mag (5$\sigma$).  The $K_s$ band mosaics fill the gap in the infrared wavelength coverage between the WFC3/IR and IRAC instruments at depths commensurate with the deep space-based imaging and with image quality that suffers significantly less from source crowding and blending than the redder IRAC bands.  

A comparison of the $K_s$-band and space-based Frontier Fields imaging is shown in Fig.~\ref{fig:cutout}, showing just a small cutout of the MACS-0416 ``parallel'' field.  The deep optical and near-infrared \textit{Hubble} imaging provides spectacular multiwavelength spatially-resolved information on physical scales of just $\sim$1~kpc (Fig.~\ref{fig:cutout}a).  However, there are many red galaxies in the indicated areas, predominantly at redshifts $z\gtrsim2$.  At $z>3$ even the reddest WFC3/IR filter, $H_{160}$ (Fig.~\ref{fig:cutout}b), only probes rest-frame ultraviolet wavelengths $<4000$~\AA\ and is therefore most sensitive to young, UV-bright star-forming galaxies.  Sampling of more evolved stellar populations at $z>3$ with redder colors requires deep imaging at longer wavelengths.  This can be obtained with imaging in the \textit{Spitzer} IRAC bands \citep[e.g.,][]{marchesini:10, stefanon:13, stefanon:15}, though at the cost of low spatial resolution ($\sim1\farcs6$, Fig.~\ref{fig:cutout}d).  This resolution is insufficient for spatially resolving the distant galaxies and is prone to blending of faint sources, particularly in the crowded Frontier Fields clusters where the surface density of both foreground cluster galaxies and faint background galaxies is high at the depths of interest.  The deep \HAWKI\ imaging described here (Fig.~\ref{fig:cutout}c) is able to bridge this gap, clearly detecting and resolving all but the faintest blue galaxies seen in the deep \textit{Hubble} images.  

Here we explore the quantitative constraints provided by the deep $K_s$ band imaging in terms of the photometric redshifts and derived intrinsic properties of galaxies in the survey field.  We construct preliminary PSF-matched photometric catalogs of the \textit{HST} imaging and deblended photometry from the IRAC bands following the methodology described by \cite{skelton:14}. We then combine the space-based catalogs with aperture photometry matched directly from $K_s$-detected catalogs derived from the completeness simulations described above\footnote{Full robust \textit{HST}+$K$+IRAC catalogs will be presented by Shipley et al. (2016), in preparation.}.  We then compute photometric redshifts with EAZY \citep{brammer:08} for the catalogs with and without the $K_s$-band photometry included.  The left panel of Fig.~\ref{fig:delta_UV} shows the difference in the rest-frame $U-V$ color derived from the photometric redshift fit, which probes the strength of the Balmer/4000~\AA\ break and is a proxy for the age and mass-to-light ratio of the underlying stellar population.  

The scatter in the $U-V$ colors with and without including the $K_s$-band information is low at $z<2$ where fit is constrained predominantly by the deep \textit{HST} photometry.  At $z>2$, however, as the rest-frame $V$ band is redshifted beyond the red $H_{160}$ filter, the scatter increases dramatically, reaching $\sigma>0.1$~mag at $z\sim3$.  This is much larger than the photometric uncertainties in the adjacent space-based photometric bands would suggest, as all of these galaxies at $H\lesssim26$ are detected in the deep WFC3 and IRAC images at $\gg10$$\sigma$.  Therefore, the large scatter is likely dominated by systematic differences in the photometric redshifts and therefore also in the derived intrinsic properties of galaxies in the survey.  

The right two panels of Fig.~\ref{fig:delta_UV} show a single galaxy that illustrates how these systematic effects are not trivial and will likely result in biases in the interpretation of the galaxy population properties derived from the \textit{HST} and IRAC observations alone.  The spectral energy distribution (SED) shown is steadily rising through the reddest WFC3/IR bands, and then shows a sharp break with bright detections in the IRAC bands.  The $K_s$-band measurement at 2.1~$\mu$m reduces the range of allowed photometric redshifts by a factor of two by pinpointing a strong Balmer break at $z\sim3.4$.  Even though the \textit{measured} $H_{160}-K_s$ color is redder than that inferred from the \textit{HST} + IRAC photometry alone, the final rest-frame $U-V$ color is actually \textit{bluer} as a result of the higher preferred redshift.  Evolved galaxies at $z>3$ such as the one shown in Fig.~\ref{fig:delta_UV}b are an intriguing population deserving of detailed study in their own right, and the combined Frontier Fields \textit{Hubble}+$K_s$+IRAC dataset is ideally suited for this purpose.

Our ultra-deep K$_{s}$-band mosaics will complement the {\it HST} and {\it Spitzer} data of the HFF, ensuring estimation of the most accurate photometric redshifts, rest-frame colors and luminosities, and stellar population properties (e.g., stellar mass and dust extinction) for galaxies at $z>2$, enhancing enormously the scientific impact of the Hubble Frontier Fields program. For example, they will allow for the investigation of the spectral energy distribution of $z\approx4$ galaxies or the evolution of the stellar mass function of galaxies at high redshift. Finally, we stress that, our mosaics more than double the total area of the sky imaged in the $K_{s}$ band down to these extreme depths. 


%


%
%

\acknowledgements

\noindent We gratefully acknowledge funding support from the STScI Director's Discretionary Research Fund. DM acknowledges the support of the Research Corporation for Science Advancement's Cottrell Scholarship, and of the National Science Foundation under Grant No. 1513473. DM and DL-V acknowledge support from program HST-AR-14302, and KEW gratefully acknowledges support by NASA through Hubble Fellowship grant \#HF2-51368, both of which are provided by NASA through a grant from the Space Telescope Science Institute, which is operated by the Association of Universities for Research in Astronomy, Incorporated, under NASA contract NAS5-26555. BV acknowledges the support from an Australian Research Council Discovery Early Career Researcher Award (PD0028506).  MT acknowledges the support by JSPS KAKENHI Grant Number JP15K17617.  This work was supported by NASA Keck PI Data Awards, administered by the NASA Exoplanet Science Institute. Some of the data presented in this study were obtained at the W. M. Keck Observatory from telescope time allocated to the National Aeronautics and Space Administration through the agency's scientific partnership with the California Institute of Technology and the University of California. The Observatory was made possible by the generous financial support of the W. M. Keck Foundation. The authors wish to recognize and acknowledge the very significant cultural role and reverence that the summit of Mauna Kea has always had within the indigenous Hawaiian community. We are most fortunate to have the opportunity to conduct observations from this mountain. This research made use of Astropy, a community-developed core Python package for Astronomy \citep{astropy}, open-source Python modules Numpy, Scipy, Matplotlib, and scikit-image, and NASA's Astrophysics Data System (ADS).  

Based on observations collected at the European Organisation for Astronomical Research in the Southern Hemisphere under ESO programmes 090.A-0458, 092.A-0472, and 095.A-0533.


\noindent {\normalsize \it Facilities:}
\facility{VLT:Yepun (HAWKI)}
\facility{Keck:I (MOSFIRE)} 

\bibliographystyle{apj_url}
\bibliography{ms}

\appendix

\label{appendix:sip}
\section{SIP keywords for \HAWKI\ geometric distortion}

The \HAWKI\ instrument has small but non-negligible geometric distortion, resulting in astrometric shifts of order 2 pixels at the corners of the detector relative to the center.  \cite{libralato:14} provide a precise determination of this distortion based on observations of star clusters.  They present a full pixel map for the distortions in $x$ and $y$ across each of the detectors.  To use this distortion model with the \textsc{drizzle} software, we fit the pixel offsets from \cite{libralato:14} with a third-order two-dimensional polynomial.  The polynomial coefficients and additional header keywords needed to create a SIP FITS distortion model \citep{fits:sip} for the four \HAWKI\ detectors are provided in Table~\ref{tab:sip}.  We adopt the \cite{libralato:14} definition of the \HAWKI\ chips, with the chip numbers set by the order they are found in the multi-extension FITS files.  That is, ``Chip \#1'' is the first image extension of the raw FITS files provided by the ESO archive, with \textsc{extname}$=$CHIP1.INT1.  The \textsc{extname} values for chips 2, 3, and 4 are CHIP2.INT1, CHIP4.INT1, and CHIP3.INT1, respectively; note the switched names of the last two chips.



\begin{deluxetable}{lrrrr}
\tabletypesize{\scriptsize}
\tablewidth{0pt} 
\tablecaption{\HAWKI\ SIP Header Keywords \label{tab:sip}} 
\tablehead{ \colhead{Keyword} & \colhead{Chip \#1\tablenotemark{a}} & \colhead{Chip \#2} & \colhead{Chip \#3} & \colhead{Chip \#4} }
\startdata 
A\_ORDER & \multicolumn{4}{c}{3} \\
B\_ORDER & \multicolumn{4}{c}{3} \\
CTYPE1 & \multicolumn{4}{c}{RA{-}{-}{-}TAN-SIP} \\
CTYPE2 & \multicolumn{4}{c}{DEC{-}{-}TAN-SIP} \\
CRPIX1 & \multicolumn{4}{c}{1024\tablenotemark{b}} \\
CRPIX2 & \multicolumn{4}{c}{1024\tablenotemark{b}} \\
A\_0\_0 &   $ 1.751\mathrm{e}-02 $ &  $  8.098\mathrm{e}-03$ &  $  9.104\mathrm{e}-04 $ &  $  9.096\mathrm{e}-03 $ \\
B\_0\_0 &   $-2.649\mathrm{e}-02 $ &  $ -9.338\mathrm{e}-03$ &  $  4.709\mathrm{e}-03 $ &  $  1.043\mathrm{e}-02 $ \\
A\_1\_0 &   $-3.924\mathrm{e}-04 $ &  $ -4.964\mathrm{e}-05$ &  $ -1.035\mathrm{e}-04 $ &  $ -6.559\mathrm{e}-05 $ \\
B\_1\_0 &   $-9.546\mathrm{e}-04 $ &  $  9.804\mathrm{e}-04$ &  $  1.228\mathrm{e}-03 $ &  $ -7.145\mathrm{e}-04 $ \\
A\_2\_0 &   $ 8.676\mathrm{e}-07 $ &  $ -7.943\mathrm{e}-07$ &  $  8.461\mathrm{e}-07 $ &  $ -8.579\mathrm{e}-07 $ \\
B\_2\_0 &   $ 2.026\mathrm{e}-07 $ &  $  2.347\mathrm{e}-07$ &  $ -1.896\mathrm{e}-07 $ &  $ -2.628\mathrm{e}-07 $ \\
A\_3\_0 &   $-2.059\mathrm{e}-10 $ &  $ -3.724\mathrm{e}-10$ &  $ -2.350\mathrm{e}-10 $ &  $ -3.941\mathrm{e}-10 $ \\
B\_3\_0 &   $-6.040\mathrm{e}-11 $ &  $  4.753\mathrm{e}-11$ &  $  1.045\mathrm{e}-12 $ &  $ -1.550\mathrm{e}-11 $ \\
A\_0\_1 &   $-4.413\mathrm{e}-05 $ &  $  6.173\mathrm{e}-05$ &  $  1.025\mathrm{e}-04 $ &  $ -5.038\mathrm{e}-05 $ \\
B\_0\_1 &   $-4.422\mathrm{e}-04 $ &  $ -5.841\mathrm{e}-04$ &  $ -1.974\mathrm{e}-04 $ &  $ -4.552\mathrm{e}-04 $ \\
A\_0\_2 &   $ 2.978\mathrm{e}-07 $ &  $ -3.050\mathrm{e}-07$ &  $  3.071\mathrm{e}-07 $ &  $ -2.771\mathrm{e}-07 $ \\
B\_0\_2 &   $ 7.265\mathrm{e}-07 $ &  $  7.453\mathrm{e}-07$ &  $ -6.841\mathrm{e}-07 $ &  $ -7.129\mathrm{e}-07 $ \\
A\_0\_3 &   $-1.277\mathrm{e}-11 $ &  $  3.299\mathrm{e}-11$ &  $ -4.466\mathrm{e}-13 $ &  $ -1.834\mathrm{e}-11 $ \\
B\_0\_3 &   $-2.177\mathrm{e}-10 $ &  $ -2.529\mathrm{e}-10$ &  $ -2.448\mathrm{e}-10 $ &  $ -2.646\mathrm{e}-10 $ \\
A\_1\_1 &   $ 6.053\mathrm{e}-07 $ &  $  6.554\mathrm{e}-07$ &  $ -3.966\mathrm{e}-07 $ &  $ -6.455\mathrm{e}-07 $ \\
B\_1\_1 &   $ 5.875\mathrm{e}-07 $ &  $ -4.337\mathrm{e}-07$ &  $  5.017\mathrm{e}-07 $ &  $ -4.399\mathrm{e}-07 $ \\
A\_1\_2 &   $-2.199\mathrm{e}-10 $ &  $ -2.919\mathrm{e}-10$ &  $ -1.881\mathrm{e}-10 $ &  $ -3.229\mathrm{e}-10 $ \\
B\_1\_2 &   $ 9.213\mathrm{e}-12 $ &  $  7.527\mathrm{e}-11$ &  $  3.588\mathrm{e}-11 $ &  $ -4.918\mathrm{e}-11 $ \\
A\_2\_1 &   $-4.114\mathrm{e}-11 $ &  $  7.218\mathrm{e}-11$ &  $ -2.636\mathrm{e}-11 $ &  $ -6.355\mathrm{e}-11 $ \\
B\_2\_1 &   $-2.118\mathrm{e}-10 $ &  $ -2.506\mathrm{e}-10$ &  $ -1.433\mathrm{e}-10 $ &  $ -2.753\mathrm{e}-10 $
\enddata
\tablenotetext{a}{We adopt the \cite{libralato:14} definition of the \HAWKI\ chips, see text.}
\tablenotetext{b}{The SIP distortion polynomial is defined relative to the reference pixel \textsc{crpix}, which we set to be the center of each detector.  This is different from the default reference pixel in the raw images so the \textsc{crval} values also have to be shifted accordingly.}
\end{deluxetable}

\end{document}